\documentstyle[a4,12pt,epsf]{article}

\newcommand{\sikib}{\begin{eqnarray}}
\newcommand{\sikie}{\end{eqnarray}}
\newcommand{\lm}{ {\mathcal{L}}_{m}}
\newcommand{\scf}{\varphi}
\newcommand{\dscf}{\dot{\varphi}}
\newcommand{\om}{\omega}
\newcommand{\ep}{\epsilon}
\newcommand{\si}{\sigma}

\newcommand{\cdmu}{\nabla_{\mu}}
\newcommand{\cdnu}{\nabla_{\nu}}
\newcommand{\efmass}{\widetilde{M}}

\title{Black holes and a scalar field in an expanding universe}

\author{ Hiromi {\sc Saida}$^a$
\footnote{E-mail: saida@phys.h.kyoto-u.ac.jp}
         and
         Jiro {\sc Soda}$^b$
\footnote{E-mail: jiro@phys.h.kyoto-u.ac.jp} \\[3mm]
  $^a$
  Graduate School of Human and Environmental Studies, 
  Kyoto University, \\ 
  Kyoto 606-8501, Japan, \\
  $^b$
  Department of Fundamental Sciences, FIHS, Kyoto University, \\
  Kyoto 606-8501, Japan }

\date{}

\begin{document}

\maketitle

\baselineskip=5.5mm

\begin{abstract}

We consider a model of inhomogeneous universe with the presence 
of a massless scalar field, where the inhomogeneity is assumed to 
consist of many black holes. This model can be constructed by 
following Lindquist and Wheeler, which has already been 
investigated without the presence of scalar field to show that an 
averaged scale factor coincides with that of the Friedmann model in 
Einstein gravity. In this paper we construct the inhomogeneous 
universe with a massless scalar field, where it is assumed that the 
averaged scale factor and scalar field are given by those of the 
Friedmann model including the scalar field. All of our calculations 
are carried out within the framework of Brans-Dicke gravity. In 
constructing the model of an inhomogeneous universe, we define the 
mass of a black hole in the Brans-Dicke expanding universe which is 
equivalent to the ADM mass in the epoch of the adiabatic time 
evolution of the mass, and obtain an equation relating our mass 
with the averaged scalar field and scale factor. As the results we 
find that the mass has an adiabatic time dependence in a 
sufficiently late stage of the expansion of the universe, that is 
our mass is equivalent to ADM mass. The other result is that its 
time dependence is qualitatively diffenrent according to the sign 
of the curvature of the universe: the mass increases in a 
decelerating fashion in the closed universe case, is constant in 
the flat case, and decreases in a decelerating fashion in  the open 
case. It is also noted that the mass in the Einstein frame depends 
on time. Our results that the mass has a time dependence should be 
retained even in the general scalar-tensor gravities with a scalar 
field potential. Furthermore, we discuss the relation of our model 
of the inhomogeneous universe with the uniqueness theorem of black 
hole spacetime and the gravitational memory effect of black holes 
in scalar-tensor gravities.

\end{abstract}

{\footnotesize PACS: 04.70.-s, 04.70.Bw, 98.80.Hw}


\section{Introduction}\label{sec-intro}

Recently strongly motivated by the supernovae observations 
\cite{ref-obs}, general agreement has been reached that our 
universe is going to turn to accelerated expansion in the context 
of Friedmann universe model. As one possible model, a scalar field 
can be introduced in the Friedmann universe \cite{ref-qe}. Such a 
scalar field plays the role as the quintessence of accelerated 
expansion. The investigation of scalar field for an expanding 
universe is an important current issue. 

On the other hand, our universe consists of celestial objects like 
galaxies, stars and possibly black holes. In order to draw a more 
precise picture of our universe than that of the Friedmann model, 
it is useful to investigate the effects of the inhomogeneities on 
the expansion of the universe
\footnote{ This point of view may lead us to study black holes in a 
non-flat background. So far only a few works related to this issue 
have been done numerically or in a toy model \cite{ref-eubh} }. 
In ref.\cite{ref-clu}, the expansion law of inhomogeneous universe 
including many black holes has already been investigated. We call 
the model of an inhomogeneous universe used in ref.\cite{ref-clu} 
the {\it cell lattice universe}. It expresses an averaged space 
time of inhomogeneous universe consisting of many identical black 
holes, which is constructed as follows: considering a regularly 
tessellated homogeneous and isotropic universe, each cell of the 
tessellation is replaced by a spherically symmetric black hole. 
Here it is assumed that the averaged values of any quantities are 
defined by the values on the junction surface where a homogeneous 
universe is connected with a spherical black hole. The behavior of 
the averaged scale factor is determined by the junction condition 
of the cell lattice universe, and reproduces the same expansion law 
that the dust-dominated Friedmann model obeys, as has already been 
shown in ref.\cite{ref-clu}. We believe that the cell lattice 
universe model gives us the well-defined averaged quantities of 
an inhomogeneous universe, and that this model coincides with the 
matter-dominated universe. 

In this paper, we extend the cell lattice universe model in order 
to include a scalar field. Motivated by ref.\cite{ref-clu}, we 
assume the averaged scalar field and scale factor are given by the 
values on the junction surface. Furthermore, we require the 
averaged quantities to agree with those of the dust-dominated 
Friedmann model including a scalar field. Here the undetermined 
quantity is the form of the metric of a spherical black hole. With 
these assumptions we are interested in the question: {\it how 
should the black hole be affected by the expansion of the universe 
and the existence of a scalar field in order to retain the 
consistency of the cell lattice universe with the dust-dominated 
Friedmann model}. That is, in contrast to ref.\cite{ref-clu} where 
the expansion law of the cell lattice universe has been 
investigated, we turn our interest to the black holes in the cell 
lattice univese. It is well known that the system of Einstein 
gravity with scalar fields can be transformed to the framework of 
scalar-tensor gravity. For simplicity, we consider the scalar field 
to be a massless one and carry out all of the calculations within 
the framework of the scalar-tensor gravity, especially of the 
Brans-Dicke gravity. In constructing the cell lattice universe in 
Brans-Dicke gravity in the way mentioned at the begining of this 
paragraph, we define the mass of the black hole by making use of 
the undetermined form of the metric. If the time dependence of our 
mass is adiabatic and the universe is expanding, the mass is 
equivalent to the ADM mass of a Schwarzschild black hole without 
regard to the details of the expansion law of the universe. The 
junction condition in replacing a cell by a black hole gives us an 
equation relating the mass to the averaged scalar field and scale 
factor. Through this equation we can investigate whether and how 
the mass evolves with time, which is an effect of the expansion of 
the universe and the existence of the scalar field on black holes. 

In sections \ref{sec-fubdg} and \ref{sec-clu}, Friedmann universe in 
Brans-Dicke gravity and the method of constructing the cell lattice 
universe are reviewed, respectively. Section \ref{sec-bhmeu} is 
devoted to the construction of the cell lattice universe in 
Brans-Dicke gravity and to analyses of the mass which is defined in 
the section \ref{sec-bhmeu}. Finally we give a summary and 
discussion in section \ref{sec-sd}. Throughout this paper, we set 
$c=1$.


\section{Friedmann Universe in Brans-Dicke Gravity}\label{sec-fubdg}

The action of Brans-Dicke gravity is
  \sikib
    S = \int d^4 x \sqrt{-g} \, 
        \left[ \scf R - \frac{\om}{\scf} (\nabla \scf)^2 
          + \lm \right] \, ,
  \label{eq-fubdg1}
  \sikie
where $\om$ is the constant parameter of this theory, $\scf$ is the 
scalar field coupling to gravity and $\lm$ is the matter Lagrangian 
which does not include $\scf$. The effective Newton ``constant'', 
$G_{eff}$, is related to the scalar field as 
$\scf = (16 \pi G_{eff})^{-1}$. The field equations derived from 
this action are 
  \sikib
    R_{\mu\nu} - \frac{1}{2} g_{\mu\nu} R
      &=& \frac{\om}{\scf^2} 
            \left[ (\cdmu \scf)(\cdnu \scf) 
             - \frac{1}{2} g_{\mu\nu} (\nabla \scf)^2 \right]
          - \frac{1}{\scf} 
            \left[ g_{\mu\nu} \Box - \cdmu \cdnu \right] \scf 
          + \frac{1}{2 \scf} T_{\mu\nu} \, , \nonumber \\
    \Box \scf &=& \frac{1}{4\om + 6} T_{\mu}^{\mu} \, ,
  \label{eq-fubdg2}
  \sikie
where $T_{\mu\nu}$ is the energy-momentum tensor derived from 
$\lm$, which is automatically divergenceless: 
$\cdmu T^{\mu\nu} = 0$. 

We consider the Friedmann universe in Brans-Dicke gravity. The 
metric is spatially homogeneous and isotropic, 
  \sikib
    ds^2 = - dt^2
           + a(t)^2 \left[ \frac{dr^2}{1 - k r^2}
           + r^2 (d\theta^2 + \sin^2 \theta d\phi^2) \right] \, ,
  \label{eq-fubdg3}
  \sikie
where $k=-1$,$0$,$1$ and $a(t)$ is the scale factor. The matter 
in the universe is of perfect fluid type,
  \sikib
    T_{\mu\nu} = \ep \, u_{\mu} u_{\nu}
               + p \, ( g_{\mu\nu} + u_{\mu} u_{\nu} ) \, ,
  \label{eq-fubdg4}
  \sikie
where $u_{\mu}$ is the 4-velocity of comoving observer, $\ep$ is 
the energy density and $p$ is the pressure. Then the field 
eqs.(\ref{eq-fubdg2}) give three independent equations,
  \sikib
    H^2 + \frac{k}{a^2}
      &=& \frac{\ep}{6\, \scf} - H \frac{\dscf}{\scf}
         + \frac{\om}{6} \left( \frac{\dscf}{\scf} \right)^2 \, ,
      \label{eq-fubdg5} \\
    \ddot{\scf} + 3 H \dscf
      &=& \frac{1}{4\om + 6} \left( \ep - 3 p \right) \, ,
      \label{eq-fubdg6} \\
    \dot{\ep} &=& -3 H \left( \ep + p \right) \, .
      \label{eq-fubdg7} \
  \sikie
where $H = \dot{a}/a$ is the Hubble parameter. This system can be 
solved provided an equation of state is specified. 

For a dust-dominated universe the equation of state is $p=0$, with 
which eq.(\ref{eq-fubdg7}) gives $\ep = \ep_0/a^3$, where 
$\ep_0$ is an integration constant. Then we obtain from 
eq.(\ref{eq-fubdg6})
  \sikib
    \scf = \frac{\ep_0}{4\om +6} \int^t dt \frac{t + t_0}{a^3} \, ,
  \label{eq-fubdg9}
  \sikie
where $t_0$ is an integration constant. Eqs.(\ref{eq-fubdg5}) and 
(\ref{eq-fubdg9}) determine the time evolution of the scale factor 
and the scalar field of the dust-dominated Friedmann universe in 
Brans-Dicke gravity. 

For a radiation-dominated universe the equation of state is 
$p=\ep/3$, then eq.(\ref{eq-fubdg7}) gives $\ep = \ep_0/a^4$, where 
$\ep_0$ is an integration constant. Using eq.(\ref{eq-fubdg6}) we 
obtain
  \sikib
    \scf = \int^t dt \frac{q}{a^3} \, ,
  \label{eq-fubdg10}
  \sikie
where $q$ is an integration constant. Eqs.(\ref{eq-fubdg5}) and 
(\ref{eq-fubdg10}) determine the time evolution of the scale factor 
and the scalar field of the radiation-dominated Friedmann universe 
in Brans-Dicke gravity.

For the use of section \ref{sec-bhmeu}, let us review the 
frame transformation. By a suitable conformal transformation of the 
metric $g_{\mu\nu}$ to the other one $\tilde{g}_{\mu\nu}$, the 
action $S$ of eq.(\ref{eq-fubdg1}) can be treated as Einstein 
gravity with a massless scalar field which does not couple to 
gravity. Such a transformation is given by 
  \sikib
    g_{\mu\nu} = 
      \exp \left[ - \frac{\si}{2\om+3} \right] \, 
        \tilde{g}_{\mu\nu} \, ,
  \label{eq-fubdg11}
  \sikie
where $\si$ is a scalar field defined by 
$\scf = (16\pi G_0)^{-1} \exp[ \, \si/(2\om+3) \, ]$. Here $G_0$ is 
the ordinary Newton constant in the Einstein gravity. Then the 
action $S$ becomes 
  \sikib
    S = \int d^4 x \sqrt{-\tilde{g}} \, 
        \left[ \frac{1}{16\pi G_0}
          \left( \tilde{R} - ( \tilde{\nabla} \si )^2 \right)
          + e^{-2\si/(2\om+3)} \lm \right] \, .
  \label{eq-fubdg12}
  \sikie
The action expressed by $\tilde{g}_{\mu\nu}$ is said to be in the 
Einstein frame, while the action of eq.(\ref{eq-fubdg1}) is in the 
original frame. We add tilde to quantities in the Einstein 
frame. It is obvious by this frame transformation that the 
Brans-Dicke gravity in the original frame becomes the Einstein 
gravity in the limit $\om \to \infty$, that is, 
$g_{\mu\nu} = \tilde{g}_{\mu\nu}$. 

It has already known analytically that, in the dust-dominated and 
flat universe case, $p=0$ and $k=0$, only the choice that $t_0=0$ 
lets the scale factor $a(t)$ in Brans-Dicke gravity go over 
smoothly to that of Einstein gravity in the limit $\om \to \infty$ 
\cite{ref-gc}. With $t_0=0$ we can find the relation
  \sikib
    \frac{8\om + 12}{3\om + 4}
      = \frac{\ep_0 \, t^2}{\scf \, a^3} \, .
  \label{eq-fubdg13}
  \sikie
The scale factor and scalar field in this case can be easily 
obtained to be 
  \sikib
    a(t) &=& a_0 \, t^{(2\om + 2)/(3\om + 4)} \, , \nonumber \\
    \scf(t) &=& \scf_0 \, t^{2/(3\om + 4)} \, , 
  \label{eq-fubdg14}
  \sikie
where $a_0$ and $\scf_0$ are constants. In the other curvature 
cases $k= \pm 1$ with a restriction $t_0=0$, eq.(\ref{eq-fubdg5}) 
indicates that the universe is dominated at early times by terms 
other than the curvature one $k/a^2$. That is, the curved universe 
is effectively flat at early times.


\section{Cell Lattice Universe}\label{sec-clu}

\subsection{Strategy to construct the cell lattice universe and 
Averaged quantities}

It is a purely geometric problem to construct the cell lattice 
universe which expresses an averaged inhomogeneous expanding 
universe including many identical spherically symmetric objects 
\cite{ref-clu}. The procedure of the construction consists of two 
steps as follows: {\it 1st step}, regularly tessellate the 
spherical, flat or hyperbolic spatial section, $\Sigma_t$, of a 
homogeneous and isotropic universe by a regular polyhedron 
\cite{ref-poly} \cite{ref-tsl}. {\it 2nd step}, replace each cell 
made of a polyhedron by a metric of the spherically symmetric 
object. For the junction condition in the 2nd step we consider the 
radially directional differential of any great circle's 
circumference on the two dimensional spherical junction surface in 
$\Sigma_t$, which can be expressed as:
  \sikib
    D = \frac{d(\mbox{Circumference of the junction surface 
              in } \Sigma_t)}{d(\mbox{Physical radial distance 
              along } \Sigma_t)} \, .
  \label{eq-clu1.0}
  \sikie
The junction condition is to connect such a differential measured 
in the homogeneous and isotropic universe, $D_{univ}$, with that 
measured in the metric of a spherically symmetric object, $D_{BH}$. 
That is the junction condition is given by 
  \sikib
    D_{univ} = D_{BH} \, .
  \label{eq-clu1.1}
  \sikie

Furthermore, in the 2nd step, there are some regions where 
spherically symmetric metrics overlap, and other regions where the 
metrics do not cover. Here we assume that the averaged values of 
any quantities are defined by the values on the spherical junction 
surface. For example, the averaged scale factor of the cell lattice 
universe is the scale factor on the spherical junction surface. The 
averaged cosmological time is set by the proper time of the 
observer staying on the spherical junction surface. We consider the 
cell lattice universe to be an averaged spacetime in such a sense. 

  \begin{figure}
    \begin{center}
      \leavevmode
      \epsfysize=55mm
      \epsfbox{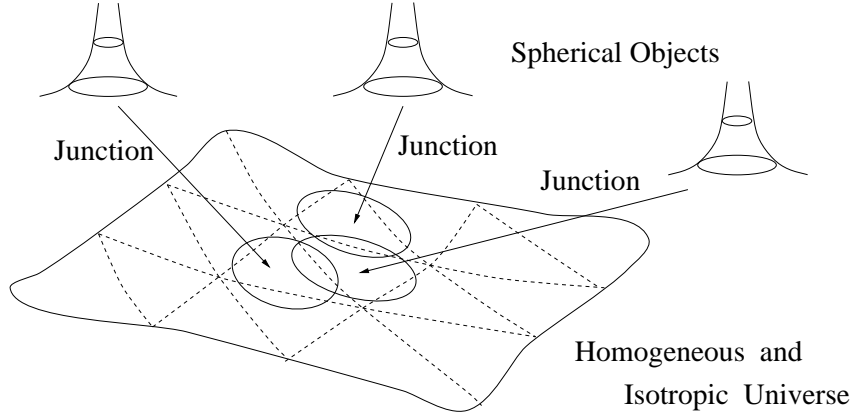}
    \end{center}
  \caption{Graphical image of the cell lattice universe. The region 
           surrounded by broken lines is the regular polyhedron. 
           The circle of real lines is the spherical cell. Each 
           cell on the homogeneous and isotropic universe is 
           replaced by a spherically symmetric object. There are 
           some region where some objects overlap, while the other 
           regions are uncovered.}
  \label{fig-clu}
  \end{figure}

\subsection{Construction of the cell lattice universe}

The metric of a homogeneous and isotropic universe can be given in 
the same form as eq.(\ref{eq-fubdg3}). The 1st step in constructing 
the cell lattice universe is to determine the radius of each 
spherical cell measured in this metric \cite{ref-poly}  
\cite{ref-tsl} by requiring that the volume of a spherical cell 
coincides with that of a regular polyhedron. There are some regular 
polyhedra which are candidates for the regular tessellation for 
each curvature case. The types of polyhedra are listed later in 
this section after the explanation of the junction condition. We 
denote the comoving radius of the spherical cell by $r_c$. Since 
the comoving number density of cells is constant, the cell expands 
comovingly. Therefore $r_c = const.$, and the averaged cosmological 
time is identical to the time coordinate of the metric 
(\ref{eq-fubdg3}), $t_c$, on the junction surface $r=r_c$. 
Hereafter we denote the physical radius of a cell by 
$l_c(t_c) = a(t_c)r_c$. 

In proceeding to the 2nd step, we define precisely the ``radial 
direction'' in the differential of the great circle on the junction 
surface. In choosing the parameter along the radial direction, 
$\chi$, into a coordinate system such as $(t, \chi, \theta, \phi)$, 
we define the coordinate parameter $\chi$ as follows: with the 
conditions that $t$, $\theta$ and $\phi$ are constant, the line 
element along the radial direction is given by 
$ds^2 = a^2 d\chi^2$, that is to say, $\chi$ is the ``comoving 
radial distance''. For convenience we transform the coordinate 
$(t, r, \theta, \phi)$ to $(t, \chi, \theta, \phi)$ by 
$r=f_k(\chi)$, where $f_k(\chi) = \sin \chi$, $\chi$ and 
$\sinh \chi$ for $k=1$, $0$ and $-1$, respectively: 
  \sikib
    ds^2 = - dt^2 + a(t)^2 \left[ \, d\chi^2
            + f_k(\chi) \, 
              (d\theta^2 + \sin \theta d\phi^2) \, \right] \, .
  \label{eq-clu0.0}
  \sikie
In denoting $r_c = f_k(\chi_c)$, this $\chi_c$ should be constant 
because $\chi_c$ is the comoving coordinate on the junction 
surface. The circumference of great circle, $C_{gc}$, is calculated 
in this metric to be 
  \sikib
    C_{gc} = 
    \int_{\Sigma_t, \, \chi=\chi_c, \, \theta=\pi/2} 
      \sqrt{g_{\phi\phi} } \, d\phi = 
    2 \pi \, a \, f_k(\chi_c) \quad ( \, = 2\pi l_c \, ) \, .
  \label{eq-clu0.1}
  \sikie
Then we obtain the radially directional differential of the great 
circle in the metric (\ref{eq-clu0.0}), $D_{univ}$, to be
  \sikib
    D_{univ}
     = \left. \frac{dC_{gc} }{d(a \chi)} 
         \right|_{\Sigma_t, \, \chi_c}
     = 2 \pi \frac{d f_k(\chi_c)}{d \chi_c} \, .
  \label{eq-clu0.2}
  \sikie

The metric of a spherically symmetric object can be given by
  \sikib
    ds^2 = - A(T,R) \, dT^2 + B(T,R) \, dR^2
           + R^2 (d\theta^2 + \sin^2 \theta d\phi^2) \, ,
  \label{eq-clu1}
  \sikie
where $A(T,R)$ and $B(T,R)$ are the arbitrary independent functions 
of $T$ and $R$. The radial coordinate $R$ is the physical radial 
length. Hereafter we assume that the radius of a spherical object 
is smaller than that of a cell, therefore the metric 
(\ref{eq-clu1}) is the exterior spacetime of the object. In the 
2nd step of constructing the cell lattice universe, we should pay 
attention to the fact that the coordinates $(T, R)$ of spherical 
object are not necessarily identical to $(t,r)$ or $(t,\chi)$ of a 
homogeneous and isotropic universe. When the origins of the 
coordinates of both metrics coincide, the radii of the junction 
surface in both coordinate systems coincide. That is, we have 
$R(t_c)=l_c(t_c)$ on the junction surface. The time $T$ of the 
metric (\ref{eq-clu1}) on the junction surface can be expressed by 
a function of $t_c$ and $r_c$ as $T_c=T(t_c, r_c)$, which should 
reflect the junction condition and the time evolution of the 
universe. The radially directional differential of the great circle 
in the metric (\ref{eq-clu1}) should be calculated on $\Sigma_t$. 
The circumference of the great circle, $C_{gc}$, is given by 
eq.(\ref{eq-clu0.1}). At the junction surface, the vector which is 
orthogonal to $\Sigma_t$ can be given by $e^{\mu}=(dT_c,dl_c,0,0)$ 
in the coordinate of metric (\ref{eq-clu1}), then the vector which 
is orthogonal to the junction surface and parallel to $\Sigma_t$ 
can be obtained to be $v^{\mu} = n (-g^{TT}dl_c, g^{RR}dT_c,0,0)$, 
where $n$ is the normalization constant. This gives us the 
difference along the radial direction as 
$d(a\chi)^2 = g_{\mu\nu}v^{\mu}v^{\nu}$. Here we determine the 
normalization, $n$, by requiring that the difference of the 
circumference be related to the vector 
$v^{\mu}$ as $dC_{gc} = 2\pi dl_c = 2\pi v^R$. For the averaged 
cosmological time, we have the relation 
$-dt_c^{\,\,2} = g_{\mu\nu}e^{\mu}e^{\nu}$. Then we can calculate 
the radially directional differential of the great circle in the 
metric (\ref{eq-clu1}), which we denote as $D_{BH}$, as follows:
  \sikib
    D_{BH} \, 
     = \, \frac{2\pi dl_c}{d(a\chi_c)}
     = \, 2\pi \, 
          \frac{g^{RR}dT_c}{ \sqrt{g_{\mu\nu}v^{\mu}v^{\nu}} } \, 
     = \, 2\pi \, 
          \sqrt{ -\frac{g_{TT} }{g_{RR} } } \frac{dT_c}{dt_c}
  \label{eq-clu0.3}
  \sikie

With the preparations done above, we proceed to calculating the 
junction condition, $D_{univ}=D_{BH}$. Substituting 
eq.(\ref{eq-clu0.2}) and the second expression of $D_{BH}$ of 
eq.(\ref{eq-clu0.3}) for the junction condition, we find the 
relation: 
$(dT_c/dl_c)^2 = \alpha_k^{\,\, 2} B/[A(\alpha_k^{\,\, 2} - 1/B)]$, 
where $\alpha_k = \cos\chi_c$, $1$ and $\cosh\chi_c$, for $k=+1$, 
$0$ and $-1$, respectively. Wiht this relation and 
$D_{univ}=D_{BH}$ with the third expression of $D_{BH}$ in 
eq.(\ref{eq-clu0.3}), we obtain the equation \cite{ref-clu}: 
  \sikib
    \left( \beta_k \, \frac{da(t_c)}{dt_c} \right)^2 = 
      \alpha_k^{\,\, 2} - \frac{1}{B(t_c,l_c)} \, ,
  \label{eq-clu2}
  \sikie
where $\beta_k = f_k(\chi_c)$. The parameters $\alpha_k$ and 
$\beta_k$ are determined by the cell radius $\chi_c$. Here we list 
the types of regular polyhedra, $\chi_c$ and $N$ in the Table 
\ref{table-polyhedra}, where $N$ is the number of the cells 
included in the spatial surface $\Sigma_t$ \cite{ref-poly} 
\cite{ref-tsl}. 

  \begin{table}
    \begin{center}
      \begin{tabular}{|c|c|c|c|} \hline
        Curvature $k$ & Polyhedron & $\chi_c$ & N \\ \hline\hline
          $+1$ & Tetrahedron  & 1.057  & 5 \\ \hline
          $+1$ & Tetrahedron  & 0.6866 & 16 \\ \hline
          $+1$ & Tetrahedron  & 0.1993 & 600 \\ \hline
          $+1$ & Cube         & 0.8832 & 8 \\ \hline
          $+1$ & Octahedron   & 0.5951 & 24 \\ \hline
          $+1$ & Dodecahedron & 0.3426 & 120 \\ \hline\hline
          $0$  & Cube & arbitrary & $\infty$ \\ \hline\hline 
          $-1$ & Cube         & 0.7185 & $\infty$ \\ \hline
          $-1$ & Dodecahedron & 1.251  & $\infty$ \\ \hline
          $-1$ & Dodecahedron & 0.9505 & $\infty$ \\ \hline
          $-1$ & Icosahedron  & 0.9747 & $\infty$ \\ \hline
      \end{tabular}
    \end{center}
  \caption{List of the polyhedra for the tessellation and the 
           radii $\chi_c$ corresponding to each polyhedron. 
           The fourth column $N$ is the number of spherical objects 
           included in the spatial surface $\Sigma_t$.
           \cite{ref-poly} \cite{ref-tsl} } 
  \label{table-polyhedra}
  \end{table}

For the cell lattice universe made using Schwarzschild black holes 
in Einstein gravity, the averaged scale factor is given by 
eq.(\ref{eq-clu2}) with $B^{-1}=1-2G_0M/R$, where $G_0$ is the 
ordinary Newton constant. 

It can be easily checked that, without regard to the polyhedron 
used for the tessellation, this equation reproduces the same 
expansion law as for a dust-dominated Friedmann universe in 
Einstein gravity \cite{ref-clu}. In this paper we assume that, even 
in any of the generalized theories of gravity, the avereged 
values of any quantities coincide with those in the dust-dominated 
Friedmann model.


\section{Time evolution of Black Holes in an Expanding Universe}
\label{sec-bhmeu}

In this section we consider the spherical object in the cell 
lattice universe to be a black hole.

\subsection{Black hole mass and cell lattice universe}

As mentioned in section \ref{sec-intro}, our interest in this 
paper is the effect of the expansion of the universe and a 
massless scalar field on black holes. We consider this issue within 
the framework of Brans-Dicke gravity and try to extract such 
effects from the cell lattice universe which is assumed to be 
meaningful as an approximation of multi-black hole spacetime even 
in Brans-Dicke gravity. This issue can be discussed by 
eq.(\ref{eq-clu2}) once the function $B(T,R)$ is specified. 

To specify the form of $B(T,R)$, we should note here that the 
uniqueness theorem \cite{ref-lssst} of the black holes in 
Brans-Dicke gravity has already been established with the 
asymptotically flat condition, not in the expanding universe 
\cite{ref-unique}. That is, the spherically symmetric non-rotating 
black hole in the asymptotically flat spacetime is of the 
Schwarzschild black hole even in Brans-Dicke gravity. Furthermore, 
it is important for specifying $B(T,R)$ to note the issue peculiar 
to the scalar-tensor gravity: {\it gravitational memory effect}. 
This issue was originally recognized in ref.\cite{ref-gme} and 
expressed as follows: when a black hole is formed in the expanding 
universe, does the scalar field on the event horizon keep the value 
at the black hole formation time during cosmological evolution? If 
the scalar field keeps its original value, it is said that the 
black hole has a gravitational memory effect, if not, the black 
hole does not have the memory
\footnote{ In ref.\cite{ref-gme}, two possible scenarios of 
cosmology are discussed: the scenario that primordial black holes 
have a gravitational memory effect, and the other scenario where 
primordial black holes don't have such effects.}. 
Concerning this issue, ref.\cite{ref-pbh} treats the perturbation 
of a stationary black hole with the boundary condition that the 
scalar field is proportional to the time in the region spatially 
far from the black hole. With this model, it is suggested in 
ref.\cite{ref-pbh} that the gravitational memory effect does not 
occur in scalar-tensor gravities. That is, the scalar field on the 
event horizon seems to have a time dependence along with 
cosmological evolution. Then we adopt a naively acceptable ansatz 
as follows: the spherically symmetric black hole composing the cell 
lattice universe in Brans-Dicke gravity is of the same form as the 
Schwarzschild black hole except for the one point that its radius 
depends on the cosmological time, $R_g(t_c)$. With this ansatz we 
define the mass of the black hole in the cell lattice universe, 
$M$, in Brans-Dicke gravity
\footnote{ There have already been some discussions concerning the 
spatial profile of the scalar field and the mass of a black hole 
and boson star in the scalar-tensor gravity, see \cite{ref-stmass} 
for examples. }
as, 
  \sikib
    M(t_c) \equiv 8 \pi \scf(t_c) R_g(t_c)\, .
  \label{eq-bhm1}
  \sikie
That is, we assume the black hole to be of ``Schwarzschild-type'' 
given by eq.(\ref{eq-clu1}) on the junction surface with
  \sikib
    A(t_c, l_c) = B(t_c, l_c)^{-1} 
                = 1 - \frac{M(t_c)}{8\pi \scf(t_c) \, l_c(t_c)} \, .
  \label{eq-bhm2}
  \sikie
This mass, $M$, certainly has the dimensions of mass, but it is not 
obvious whether this definition can be consistent with the ADM mass 
defined by the generator of Killing time translation in a 
stationary space time. Without regard to the details of the 
expansion law of the averaged scale factor, provided that $M$ 
changes adiabatically over a cosmological time scale and that the 
black hole horizon size is much smaller than the cosmological one, 
we can consider that the black hole exists in a local 
asymptotically flat region as for an ordinary Schwarzschild black 
hole and that the cosmological time $t_c$ is equivalent to the time 
$T$ of the metric (\ref{eq-clu1}) within the scale of the black 
hole. In such a case our mass $M$ can be effectively treated as the 
ADM mass defined by the generator of the local asymptotically 
timelike Killing vector, where the Killing time is $T$ of the 
matric (\ref{eq-clu1}). 

With specifying $B(T,R)$ as above, the junction condition 
(\ref{eq-clu2}) gives 
  \sikib
    M(t_c) = 8 \pi \beta_k^{\,\, 3} \, \scf(t_c) \, a(t_c)
            \left[ \, \dot{a}(t_c)^2 + k \, \right] \, .
  \label{eq-bhm3}
  \sikie
Clearly, since $\beta_k$ is only related to the normalization of 
the mass, the behavior of $M(t_c)$ is not affected by the value of 
$\chi_c$, that is, it is not affeced by the choice of the 
polyhedron for the tessellation in the 1st step of the construction 
of the cell lattice universe. Here, motivated by the comment in the 
final paragraph in the previous section, we assume the cell lattice 
universe to be a well-defined averaged model of an inhomogeneous 
universe, reproducing the expansion law of a dust-dominated 
Friedmann model even in Brans-Dicke gravity. Because of this 
assumption, the averaged scale factor, $a(t_c)$, and the scalar 
field, $\scf(t_c)$, are given by the dust-dominated Friedmann 
universe in Brans-Dicke gravity. The time evolution of $M(t_c)$ can 
be investigated using eqs.(\ref{eq-fubdg5}), (\ref{eq-fubdg9}) and 
(\ref{eq-bhm3}). It is not {\it a priori} obvious whether or not 
the mass $M$ depends adiabatically on the cosmological time. 

As has already been pointed out in ref.\cite{ref-pbh}, provided 
our mass $M$ can be considered as the ADM mass in a local 
asymptotically flat region, we can discuss whether the mass in the 
Einstein frame changes with time. Attaching a tilde to the quantity 
in the Einstein frame, we find the relation 
$M/\efmass = d\widetilde{T}/dT$, because the ADM mass is given by 
the generator of the local asymptotic Killing vector, which is 
normalized at the boundary of the local asymptotically flat region 
and scales inversely with the time $T$. Since the Killing time 
translation is given by the line element along integral curves of 
the timelike Killing vector, we obtain the relation at the 
boundary of the local asymptotically flat region: 
$-dT^2 = -\exp[-\si/(2\om+3)] \, d\widetilde{T}^2$, where $dT$ and 
$d\widetilde{T}$ express the Killing time translations in the 
Brans-Dicke and the Einstein frames, respectively, and the frame 
transformation given by eq.(\ref{eq-fubdg11}) is used. Therefore 
we obtain the relation of the masses \cite{ref-pbh}
  \sikib
    M = \exp \left[ \frac{\si}{4\om+6} \right] \, \efmass 
    \propto \sqrt{\scf} \, \efmass \, .
  \label{eq-bhm4}
  \sikie
If $\efmass$ is constant, $M$ is proportional to $\sqrt{\scf}$. In 
the case that the $M$ defined by ep.(\ref{eq-bhm3}) can be treated 
as the ADM mass, by comparing the $M$ with $\sqrt{\scf}$, we can 
obtain a suggestion concerning the time dependence of $\efmass$ 
which is the black hole mass in an expanding universe in Einstein 
gravity with the presence of a scalar field.

\subsection{Analysis of the black hole mass}

In the numerical calculations shown here, our interests are the 
time dependence of $M$, not the absolute value of it. That is, we 
investigate whether or not $M$ changes adiabatically on a 
cosmological time scale and whether $M$ is proportional to 
$\sqrt{\scf}$ or not, since these issues reveal that our black hole 
mass in the Brans-Dicke frame is equivalent to the ADM mass and 
that the black hole mass in the Einstein frame has a time 
dependence. Hereafter we denote the averaged cosmological time 
$t_c$ simply by $t$.

In flat case $k=0$, it is easily found from eqs.(\ref{eq-fubdg14}) 
and (\ref{eq-bhm3}) that the mass $M$ is constant, and that the 
radius $R_g(t) \propto 1/\scf(t)$ decreases in a decelerating 
fashion, while the averaged scale factor continues to expand. 
Therefore, the mass can be effectively treated as ADM mass after 
the universe expands enough to be much larger than the black hole. 
Clearly $M \not\propto \sqrt{\scf}$. This means that the mass in 
the Einstein frame has a time dependence. 

In two cases $k=\pm 1$, we calculate the mass $M(t)$ numerically 
using {\it Mathematica} from $t=0.5$ to $t=50000$ for the open case 
$k=-1$, and to $t=1000$ for the closed case $k=1$. We attach the 
subscript ``$i$'' to the quantities evaluated at $t=0.5$, and the 
subscript ``$f$'' at $t=50000$ for $k=-1$ and at $t=1000$ for 
$k=1$. These calculation time intervals of calculation should 
correspond to the dust-dominated era of our universe, since the 
cell lattice universe is assumed to be the averaged spacetime of a 
dust-dominated inhomogeneous universe as discussed in the previous 
subsection. We regard the end time of the calculation as our 
present time, because, as shown in the following, the rate of 
change of the mass $M$ in time evolution is sufficiently small in 
comparison with the averaged Hubble parameter at the end time for 
both cases $k=\pm 1$, that is $M$ evolves adiabatically. Here the 
averaged Hubble parameter $H=\dot{a}/a$ is defined by using the 
averaged scale factor. As mentioned at the end of section 
\ref{sec-fubdg}, the curved Friedmann universe can be effectively 
treated as being flat in the early stage of the time evolution. 
Therefore, we set the integration constant $t_0$ of 
eq.(\ref{eq-fubdg9}) to zero ($t_0 =0$) in order to reproduce the 
expansion law of the Friedmann model in Einstein gravity as 
$\om \to \infty$. The initial values of the averaged scale factor 
$a_i$ and scalar field $\scf_i$ are related through 
eq.(\ref{eq-fubdg13}). Our choices of the other parameters are as 
follows: the initial value of scale factor, $a_i=10$. The 
Brans-Dicke parameter $\om =500$, which is the experimental lower 
bound of $\om$. Another integration constant $\ep_0 =100$. The size 
of one cell in the cell lattice universe $\chi_c= 0.9747$ 
(Icosahedron) and $0.1993$ (Tetrahedron) for $k=-1$ and $1$, 
respectively \cite{ref-poly} \cite{ref-tsl}. The value of $\chi_c$ 
does not affect the behavior of $M$ as shown in eq.(\ref{eq-bhm3}). 

\begin{figure}
    \begin{center}
      \leavevmode
      \epsfysize=55mm
      \epsfbox{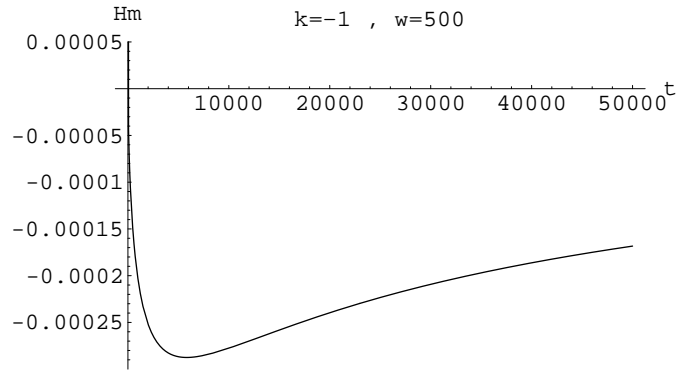}
    \end{center}
  \caption{Graph of $H_m = (\dot{M}/M)/H$ for $k=-1$, the change 
           rate of the mass. $H_{m \, i}=1.00283$, 
           $H_m(2) = 0.0001469$ and $H_m(14.92)=0$.}
  \label{fig-open.obs}
  \end{figure}

\begin{figure}
    \begin{center}
      \leavevmode
      \epsfysize=55mm
      \epsfbox{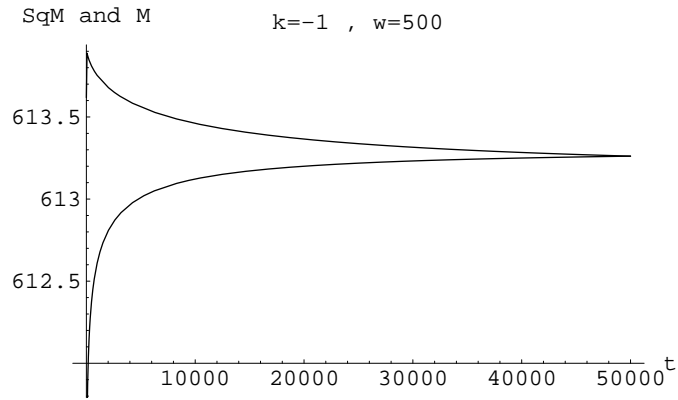}
    \end{center}
  \caption{Graph of $M(t)$ and $SqM(t)$ for $k=-1$. The upper curve 
           is for $M$ while the bottom one for $SqM$. In this graph 
           the overall scale of $SqM$ is set so that 
           $M_f=SqM_f$. $M_i=613.6$, $M(14.29)=613.9$ and 
           $SqM_i=609.4$.}
  \label{fig-open.mass}
  \end{figure}

For the open case $k=-1$, the averaged Hubble time at present 
$t=50000$ is $1/H_f = 53931$. The table \ref{table-open} shows some 
values of the averaged cosmological red shift which is defined by 
$z(t) = a_f/a(t) - 1$ with the averaged scale factor $a$. 
Fig.\ref{fig-open.obs} is for the rate of change of $M(t)$ 
normalized by the averaged Hubble parameter: 
$H_m(t) = (\dot{M}/M)/H$. The $H_m$ is of negative and asymptotes 
to zero for sufficiently late stage of cosmological evolution 
$t>10000$ where $z<3.170$. The rate of change, for example at 
$t=2$, takes the value $H_m(2)=0.0001469$. This indicates that the 
time evolution of $M$ is less significant than that of the averaged 
scale factor at least for $t>2 \,\,\, (z<1840)$, where 
$H_m < O(10^{-4})$. That is, $M$ is adiabatic in this epoch. The 
radius of black hole $R_g$ decreases in a decelerating fashion with 
time, from $R_{g \, i}=5210$ to $R_{g \, f}=5142$, while the 
averaged scale factor continues to expand for all time. Therefore, 
the mass $M$ can be effectively treated as being ADM mass after the 
universe expands enough to be much larger than the black hole. 
Fig.\ref{fig-open.mass} is the plot of $M$ with $SqM$ which is 
defined by $SqM(t) = M_f \sqrt{\scf(t)/\scf_f}$. The square root of 
the averaged scalar field $SqM$ at the initial time is 
$SqM_i=609.4$, and increases in a decelerating fashion for all 
time. The rate of change of mass $H_m$ becomes zero at $t=14.92$. 
The mass $M$ increases from $M_i=613.6$ to $M(14.29)=613.9$, and 
decreases in a decelerating fashion after the time $t=14.92$ where 
$z(14.92)=17.25$. Fig.\ref{fig-open.mass} indicates that mass $M$ 
does not coincides with $SqM$, and that the mass in the Einstein 
frame has a time dependence. 

  \begin{table}
    \begin{center}
      \begin{tabular}{|c||c|c|c|c|c|c|c|c|c|c|} \hline
        Time $t$ & 0.5 & 2 & 10 & 20 & 100 & 500 & 2000 & 5000
          & 10000 & 30000 \\ \hline
        Red Shift $z$
          & 5559 & 1840 & 601.8 & 375.8 & 125.5 & 40.42 & 14.06
          & 6.376 & 3.170 & 0.5964 \\ \hline
      \end{tabular}
    \end{center}
  \caption{Table of the averaged cosmological time and red shift
           for the open case $k=-1$. The present time is $t=50000$.} 
  \label{table-open}
  \end{table}

\begin{figure}
    \begin{center}
      \leavevmode
      \epsfysize=55mm
      \epsfbox{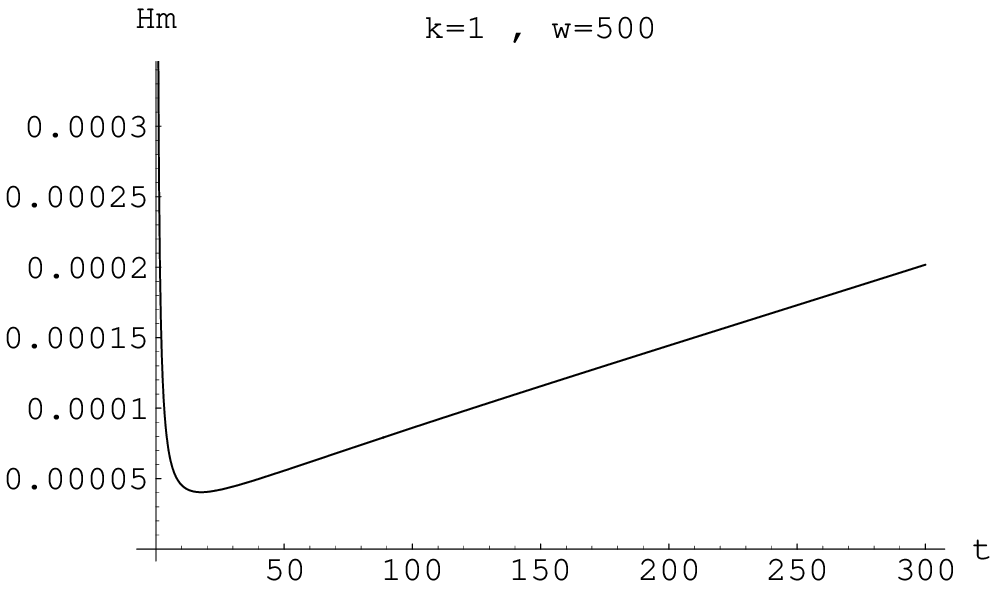}
    \end{center}
  \caption{Graph of $H_m = (\dot{M}/M)/H$ for $k=1$, the change 
           rate of the mass. $H_{m \, i}=1.00282$, 
           $H_m(2)=0.0001565$ and $H_{m \, f}=0.0006698$.}
  \label{fig-closed.obs}
  \end{figure}

\begin{figure}
    \begin{center}
      \leavevmode
      \epsfysize=55mm
      \epsfbox{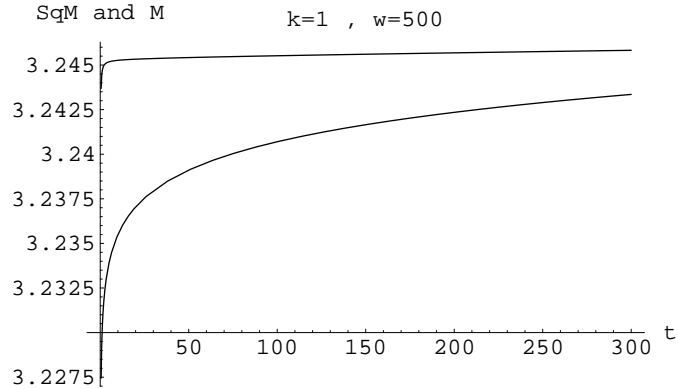}
    \end{center}
  \caption{Graph of $M(t)$ and $SqM(t)$ for $k=1$. The upper curve 
          is for $M$ while the bottom one for $SqM$. In this graph 
           the overall scale of $SqM$ is set so that 
           $M_f=SqM_f$. $M_i=3.244$ and $SqM_i=3.227$.}
  \label{fig-closed.mass}
  \end{figure}

For the closed case $k=1$, the averaged Hubble time at present 
$t=1000$ is $1/H_f = 1743$. Table \ref{table-closed} includes some 
values of the averaged cosmological red shift $z(t) = a_f/a(t) - 1$. 
Fig.\ref{fig-closed.obs} shows $H_m$, the rate of change of $M$ 
normalized by the averaged Hubble parameter. The rate of change 
$H_m$ decreases rapidly at early times, and turns to increase 
around $t=20$ where $z=11.09$. At present, the rate of change is 
$H_{m \, f} = 0.0006698$. The increase of $H_m$ continues up to the 
turning time of the averaged scale factor from expansion to 
contraction, where the turning time is $t=5483$. The rate of 
change, for example at $t=2$, is $H_m(2)=0.0001565$. This indicates 
that the time evolution of the mass $M$ is adiabatic within the 
epoch from $t=2$ to the present $t=1000$, where $H_m < O(10^{-4})$. 
The black hole radius $R_g$ decreases in a decelerating fashion 
with time, from $R_{g \, i}=27.54$ to $R_{g \, f}=27.29$, while the 
averaged scale factor lasts to expand untill the turning time 
$t=5483$. Therefore, the mass $M$ can be effectively treated as ADM 
mass after the universe expands to a size much larger than the 
black hole. Fig.\ref{fig-closed.mass} shows a plot of $M$ and 
$SqM$. The square root of the averaged scalar field $SqM$ increases 
in a decelerating fashion for all time from the initial 
value $SqM_i=3.227$. The mass $M$ takes the initial value 
$M_i=3.224$, and also increases in a decelerating fashion for all 
time. Fig.\ref{fig-closed.mass} indicates that $M$ and $SqM$ do not 
coincide with each other, and that the mass in the Einstein frame 
has a time dependence.

  \begin{table}
    \begin{center}
      \begin{tabular}{|c||c|c|c|c|c|c|c|c|c|c|} \hline
        Time $t$ & 0.5 & 2 & 5 & 10 & 15 & 20 & 50 & 100
                 & 300 & 500\\ \hline
        Red Shift $z$
          & 174.4 & 57.27 & 29.73 & 18.21 & 13.64 & 11.09
          & 5.592 & 3.189 & 1.072 & 0.5085 \\ \hline
      \end{tabular}
    \end{center}
  \caption{Table of the averaged cosmological time and red shift
           for the closed case $k=1$. The present time is 
           $t=1000$.} 
  \label{table-closed}
  \end{table}

\section{Summary and Discussion}\label{sec-sd}

In order to investigate the effects of expansion of the universe 
and a scalar field on the celestial objects which compose the 
inhomogeneity of the universe, we constructed the cell lattice 
universe in Brans-Dicke gravity. The assumptions in constructing 
the universe were that the black holes included in the universe are 
of ``Schwarzschild-type'', eqs.(\ref{eq-clu1}) and 
(\ref{eq-bhm2}), and that the averaged scale factor and scalar 
field are given by those of the Friedmann universe. Furthermore, we 
defined the mass $M$ by eq.(\ref{eq-bhm1}) which is equivalent to 
the ordinary ADM mass in the case of adiabatic time evolution of 
$M$. The junction condition of the cell lattice universe gives 
eq.(\ref{eq-bhm3}) to calculate the time evolution of the black 
hole mass $M$. It is not {\it a priori} known whether and how the 
mass depends on time. 

As a result, it turns out that $M$ behaves in a qualitatively 
different way with respect to the value of $k$. The mass $M$ 
decreases in a decelerating fashion for the open case $k=-1$, stays 
completely constant for the flat case $k=0$ and increases in a 
decelerating fashion for the closed case $k=1$. The rate of change 
of the mass $H_m$ is very small for both cases of $k=\pm 1$, 
$H_m < O(10^{-4})$ with our numerical results. This means that $M$ 
evolves adiabatically in time. Because here we assume that the 
averaged scale factor in eq.(\ref{eq-bhm3}) is given by that of the 
Friedmann model in Brans-Dicke gravity, the adiabaticity of $M$ is 
consistent with the fact that the cell lattice universe constructed 
with the Schwarzschild black hole, whose mass is completely 
constant, reproduces the expansion law of the Friedmann universe in 
Einstein gravity \cite{ref-clu}. Furthermore, the radius $R_g$ 
decreases in a decelerating fashion for every case of $k=\pm 1$, 
$0$, while the averaged scale factor continues to increase. This 
behavior of $R_g$ can be easily understood for the case $k=0$, 
which is $R_g \propto t^{-2/(3\om + 4)}$. Therefore, it is 
reasonable to consider the black hole is in a local asymptotically 
flat region after the universe expands enough to be much larger 
than the black hole, consequently mass $M$ is equivalent to the ADM 
mass defined by the local asymptotic Killing time translation. 

Furthermore, we can recognize using eq.(\ref{eq-bhm4}) that the 
black hole mass in the Einstein frame has a time dependence for 
every case of $k=\pm 1$, $0$. According to the uniqueness theorem 
\cite{ref-lssst} \cite{ref-unique}, non-rotating and non-charged 
black hole in asymptotically flat spacetime is specified only by 
the ADM mass, which is completely constant in this theorem. The 
time dependence of the mass may indicate that the uniqueness 
theorem in an expanding universe will be broken on a cosmological 
time scale. It is a very interesting problem which remains to be 
solved. 

In the numerical calculations shown in the section \ref{sec-bhmeu}, 
we set $\om=500$ which is the lowest experimental value. If the 
parameter is set as $\om<500$, the absolute value of the rate of 
change of the mass $M$ tends to increase. However the total 
behavior of the mass $M$ and the radius $R_g$ for the case 
$\om<500$ are the same as for the case $\om=500$. Consequently, for 
arbitrary value of $\om$, the mass can be equivalent to the ADM 
mass once the universe expands enough to be much larger than the 
black hole size. 

So far we have only treated the Brans-Dicke gravity as a 
representative case of scalar-tensor gravities in which the 
parameter $\om$ depends on the scalar field, and have not 
introduced any potential of the scalar field. When, instead of 
Brans-Dicke gravity, we consider a general scalar-tensor gravity 
with a potential of the scalar field in constructing the cell 
lattice universe, the averaged scale factor and scalar field may 
behave quite differently from those in Brans-Dicke gravity. 
However, the same arguments and results obtained until the previous 
paragraph are true of this case if the averaged quantities 
asymptote to those in the Brans-Dicke gravity at least in a 
sufficiently late stage of the expansion of the universe. 
Furthermore, in the case that the averaged scale factor or scalar 
field in a general theory of gravity always behaves differently 
from that in the Brans-Dicke gravity, though we need to reanalyze 
eq.(\ref{eq-bhm3}) in order to know the details of the time 
dependence of $M$, however, it is natural to propose that the mass 
$M$ depends on time in this case too. That is, it seems to be quite 
a general point that the black hole mass in expanding universe with 
the presence of scalar field has some time dependence. 

Let us comment on our assumption, i.e. the ``Schwarzschild-type'' 
ansatz of the black hole composing the cell lattice universe given 
by eqs.(\ref{eq-clu1}), (\ref{eq-bhm1}) and (\ref{eq-bhm2}). The 
construction of the cell lattice universe where a cell is replaced 
by a black hole, means that the dust-matter on homogeneous and 
isotropic universe in a cell is concentrated at a center point of 
the cell. Therefore, it is appropriate to consider the black hole 
spacetime replacing the cell includes only the scalar field 
coupling with gravity in the Brans-Dicke frame. The field equation 
of such a scalar field is obtained by substituting $T_{\mu\nu}=0$ 
into eqs.(\ref{eq-fubdg2}), 
  \sikib
    G_{\mu\nu} = 
      \frac{\om}{\scf^2} \left[ (\cdmu \scf)(\cdnu \scf) 
             - \frac{1}{2} g_{\mu\nu} (\nabla \scf)^2 \right]
    + \frac{1}{\scf} \, \cdmu \cdnu \scf \, ,
  \label{eq-sd1}
  \sikie
where the non-zero components of the Einstein tensor are calculated 
from eqs.(\ref{eq-clu1}), (\ref{eq-bhm1}) and (\ref{eq-bhm2}), 
  \sikib
    G_{01} &=& \frac{\dot{R_g} }{R^2 - R \cdot R_g } \, ,
  \nonumber \\
    G_{22} &=& - R^3 \left[ \frac{\ddot{R_g} }{2 (R - R_g)^2}
                   + \frac{\dot{R_g}^2 }{(R - R_g)^3} \right] \, ,
  \label{eq-sd2} \\
    G_{33} &=& \sin^2 \theta \, G_{22}
  \nonumber \, ,
  \sikie
where $\dot{R_g}=dR_g/dT$ with the time $T$ of metric 
(\ref{eq-clu1}). The scalar field on the Schwarzschild-type 
spacetime should satisfy the above field eqs.(\ref{eq-sd1}) and 
(\ref{eq-sd2}). 

Our ansatz of the Schwarzschild-type black hole is motivated by the 
uniqueness theorem of the black hole in asymptotically flat 
spacetime \cite{ref-unique} and the indication in 
ref.\cite{ref-pbh} that the gravitational memory effect seems not 
to occur. Conversely, in paying attention to the latter motivation, 
the gravitational memory effect can be discussed using 
eqs.(\ref{eq-sd1}) and (\ref{eq-sd2}) with the boundary condition 
$\scf(T)$ at $R=l_c(T)$ given by the scalar field of Friedmann 
model, where $l_c(T)$ is determined by the junction of our cell 
lattice universe. By investigating this system we can know whether 
or not the scalar field $\scf$ has a spatial dependence. If not, it 
means that the graviational memory effect does not occur in 
Brans-Dicke gravity, and that the indication in ref.\cite{ref-pbh} 
and our discussion in this paper are supported. The model of 
spacetime in such an approach to the gravitational memory effect, 
can be considered as a modified {\it swiss cheese universe} 
\cite{ref-swiss}. In the ordinary swiss cheese model, a spherically 
symmetric region in the Friedmann universe is replaced by an 
ordinary Shcwarzschild black hole of constant radius and mass, but 
here we replace the spherical region by a ``Schwarzschild-type'' 
black hole. This is an interesting and solvable problem.


\section*{Acknowledgments}

We would like to Thank M.Sakagami and A.Ohashi for their useful 
comments and discussions.



\begin{thebibliography}{99}

\bibitem{ref-obs}
  A.G.Riess et al, A.J.{\bf 116}(1998)1009.

\bibitem{ref-qe}
  P.Bin\'{e}truy, "Lectures at Les Houches Summer School", 
    hep-ph/0005037; 
  L.Wang, R.R.Caldwell, J.P.Ostriker and P.J.Steinhardt, 
    Ap.J.{\bf 530}(2000)17.

\bibitem{ref-eubh}
  See for examples: 
  M.Shibata and M.Sasaki, Phys.Rev.D{\bf 60}(1999)084002;
  K.R.Nayak, M.A.H.Mac Callum and C.V.Vishveshwara, gr-qc/0006040.

\bibitem{ref-clu}
  R.W.Lindquist and J.A.Wheeler, Rev.Mod.Phys.{\bf 29}(1957)432.

\bibitem{ref-gc}
  See chap.16 and the references therein: S.Weinberg, 
    ``Gravitation and cosmology'', 
    John Wiley \& Sons, Inc.(1972).

\bibitem{ref-poly}
  D.M.Y.Sommerville,
     "An Intorduction to the Geometry of n Dimensions",
     Methuen London(1929);
  H.S.M.Coxeter, "Regular Plytopos", Dover New York(1973).

\bibitem{ref-tsl}
  See sec.2 and references therein: I.H.Redmount, 
  Mon.Not.R.astr.Soc.{\bf 235}(1988)1301.

\bibitem{ref-lssst}
  See the following and the references therein: 
  S.W.Hawking and G.F.R.Ellis, 
    ``The large scale structure of space-time'', 
    Cambridge Univ.Press(1973).

\bibitem{ref-unique}
  S.W.Hawking, Commun.Math.Phys.{\bf 25}(1972)167.

\bibitem{ref-gme}
  J.D.Barrow, Phys.Rev.D{\bf 46}(1992)R3227; 
  J.D.Barrow and B.J.Carr, Phys.Rev.D{\bf 54}(1996)3920.

\bibitem{ref-pbh}
  T.Jacobson, Phys.Rev.Lett.{\bf 83}(1999)2699.

\bibitem{ref-stmass}
  See for expamples: 
    D.L.Lee, Phys.Rev.D{\bf 10}(1974)2347;
    T.Damour and G.Esposito-Frarese,
        Phys.Rev.Lett.{\bf 70}(1993)2220;
    G.L.Comer and H.Shinkai, Class.Quant.Grav.{\bf 15}(1998)669;
    A.W.Whinnett, Phys.Rev.D{\bf 61}(2000)124014

\bibitem{ref-swiss}
  C.C.Dyer and C.Oliwa, astro-ph/0004090.

\end{thebibliography}
\end{document}